\documentclass[aps,prb,twocolumn,showpacs,superscriptaddress]{revtex4-2}  
\usepackage{graphicx}  
\usepackage{dcolumn}   
\usepackage{bm}        
\usepackage{amssymb}   
\usepackage{amsmath}   
\usepackage{subfigure}   
\usepackage{chemformula} 
\usepackage{hyperref}
\hypersetup{
    colorlinks=true,
    linkcolor=blue,
    citecolor=blue,      
    urlcolor=blue,
    pdftitle={Navigating the Ti-C-O and Al-C-O ternary systems through theory-driven discovery},
    bookmarks=true,
}

\begin{document}

\title{Navigating the Ti-C-O and Al-C-O ternary systems through theory-driven discovery} 

\author{Joseph R.~Nelson} \email{jn336@cam.ac.uk} \affiliation{Department of Materials Science
  and Metallurgy, University of Cambridge, 27 Charles Babbage Road,
  Cambridge CB3 0FS, United Kingdom} \affiliation{Advanced Institute
  for Materials Research, Tohoku University, 2-1-1 Katahira, Aoba,
  Sendai, 980-8577, Japan}

\author{Richard J.~Needs} \affiliation{Theory of Condensed Matter Group,
  Cavendish Laboratory, J.~J.~Thomson Avenue, Cambridge CB3 0HE,
  United Kingdom}

\author{Chris J.~Pickard} \affiliation{Department of Materials Science and Metallurgy, University of Cambridge, 27 Charles Babbage Road,
  Cambridge CB3 0FS, United Kingdom} \affiliation{Advanced Institute
  for Materials Research, Tohoku University, 2-1-1 Katahira, Aoba,
  Sendai, 980-8577, Japan} \vskip 0.25cm

\date{\today}

\begin{abstract}
Computational searches for new materials are naturally turning from binary systems, to ternary and other multicomponent systems, and beyond. Here, we select the industrially-relevant metals titanium and aluminium and report the results of an extensive structure prediction study on the ternary titanium-carbon-oxygen \mbox{(Ti-C-O)} and aluminium-carbon-oxygen \mbox{(Al-C-O)} systems. We map out for the first time the full phase stability of \mbox{Ti-C-O} and \mbox{Al-C-O} compounds using first-principles calculations, through simple, efficient and highly parallel random structure searching in conjunction with techniques based on complex network theory. These phase stabilities emerge naturally from our `data agnostic' approach, in which we map stable compounds without recourse to structural databases or other prior knowledge. A surprising find is the predicted ambient pressure stability of octet-rule-fulfiling titanium and aluminium carbonate: \ch{Ti(CO3)2} and \ch{Al2(CO3)3}, neither of which has to our knowledge been synthesised before. These materials could potentially act as carbon sequestering compounds. Our searches discover several additional stable and metastable ternary compounds supported by the Ti-C-O and Al-C-O systems.
\end{abstract}

\pacs{}

\maketitle

\section{Introduction}   
Titanium and aluminium are two lightweight and low-density metals of tremendous industrial significance. Both are extracted from their oxide ores $-$ titania (\ch{TiO2}) and alumina (\ch{Al2O3}) $-$ with worldwide production of these raw materials exceeding a combined total of 130 million tons in 2020 \cite{USGS-1,USGS-2}. In addition to oxides, titanium and aluminium form carbides: the hard ceramic titanium carbide (\ch{TiC}), and aluminium carbide (\ch{Al4C3}). 

Bulk titania and alumina, and surfaces thereof, find several applications in catalysis and photovoltaics. Anatase-TiO$_2$ is a well-known battery electrode material \cite{deKlerk_PRM_2017}, and microporous titania \cite{Ma_AEM_2018} has also been identified as a potential sodium-ion battery anode material \cite{Choe_PCCP}, as have monolayer titanium carbides such as \ch{TiC3} \cite{Yu_JACS_2018}. Transition metal oxides show excellent potential as Li-ion battery components \cite{Harper_JM_2020}, and can offer high-rate intercalation, such as is the case with niobium tungsten oxides \cite{Griffith_Nature_2018,Kocer_ChemMater_2020}. Other so-called `MXene' compounds, consisting of two-dimensional titanium carbides with various termination groups on the surface, have been investigated for perovskite solar cells \cite{Agresti-NMat-2019}. Nanosheets, nanoparticles and surfaces of titania are reactive with \ch{CO2}, where photocatalytic reduction of \ch{CO2} is possible \cite{He_EnergyFuels_2014,Rossetti_CST_2015,Mishra_PRM_2018}, as is the adsorption of \ch{CO} and \ch{CO2} on \ch{TiO2} surfaces \cite{Reticcioli_PRL_2019,Tanaka_JPC_1984}. Titania and alumina surfaces are also under consideration as catalysts for water splitting \cite{Lei_PRM_2020,Harmon_PRM_2020}. As a result of this remarkable diversity of chemistries and material applications, the structure and stability of binary Ti-O, Ti-C, Al-O and Al-C compounds has received significant attention.

Computational structure prediction has emerged as a powerful tool in materials discovery \cite{Oganov-NatRevMat-2019}, and the accurate computational prediction of the structures of crystalline elemental or binary compounds is now, in many cases, routine. In the quest for new materials, attention has turned toward ternary or quarternary systems, in search of wider design spaces, and ever more favourable electronic or bonding properties. Ternaries are common among the known transparent conducting oxides \cite{Brunin-CompMat-2019}, and numerous examples have more recently been found in the known inorganic nitrides \cite{Sun-NMat-2019}. The larger chemical space spanned by ternary systems presents a challenge, in the form of additional complexity, to structure prediction. Materials informatics, machine learning, sensible chemical substitutions into known compounds and structures \cite{Sun-NMat-2019}, derived from databases like the Inorganic Crystal Structure Database (ICSD) \cite {ICSD} or the Open Quantum Materials Database (OQMD) \cite{OQMD}, offer some predictive power and insight into ternary systems. Other research efforts along these lines include AFLOWLIB \cite{AFLOWLIB} and the Materials Project \cite{MaterialsProject}.

Here, we explore the full Ti-C-O and Al-C-O ternary systems, determining the stable compounds present in each. We categorise our approach as \textit{theory-driven}, in contrast to data-driven, in the sense that we do not require prior knowledge to begin our searches. We use the random structure search technique \cite{Pickard-NMat-2010,Pickard-PRL-2006,Pickard-JPCM-2011}, coupled with powerful techniques inspired by complex networks which permit the automatic decomposition of crystal structures into modules or `units' \cite{Ahnert-CompMat-2017}. This grouping of atoms into units constitutes a kind of coarse-graining, which reduces the size of the configuration space to be searched, and its complexity. Inspired by the preceeding discussion on titanium and aluminium oxides and carbides and their technological importance, we chose the \mbox{Ti-C-O} and \mbox{Al-C-O} design spaces at ambient pressure \mbox{(`0 GPa')} to showcase our method. Other combinations, such as \mbox{Al-Ti-O}, are of course also possible \cite{Das_JPE_2002}. 

Some smaller-scale explorations have been made into ternary compounds containing Al, Ti, C or O and combinations thereof. Titanium oxycarbides of the form \ch{TiC_{$x$}O_{1-$x$}} can be successfully synthesized at ambient pressures \cite{Miller_JMCA_2016} by sintering titanium monoxide (\ch{TiO}) and \ch{TiC}. These materials have tunable electronic properties \cite{Brand_JPCC_2015}, and have been suggested for use in solid-oxide fuel cells (SOFCs) \cite{Sinha_PP}. Select parts of the Ti-C-O design space have been explored previously through computational means; for example the \ch{TiO2}$-$\ch{TiC} pseudobinary, \textit{viz}.~the set of compounds with chemical formulae \ch{(TiO2)_{$x$}(TiC)_{$y$}}, has been searched \cite{Meng_SciRep_2014}, with \ch{Ti5C2O6} predicted to become stable at 50 GPa. Stable ambient pressure compounds were not encountered. A restriction to pseudobinary compounds is not a requirement of our approach here, and indeed as we demonstrate below, the \mbox{Ti-C-O} system supports many compounds that do not lie along pseudobinary lines.

The rest of this paper is organised as follows. In Sec.~\ref{sec:Methods}, we give our Methods, detailing the first-principles and structure prediction calculations used in this work. In Sec.~\ref{sec:Results} we give our Results, starting with the Ti-C-O ternary system before moving on to the Al-C-O system. In Sec.~\ref{sec:Discussion} we give a discussion of our results, before concluding with Sec.~\ref{sec:Conclusions}.

\begin{figure*}
\centering
\includegraphics[width=17.0cm]{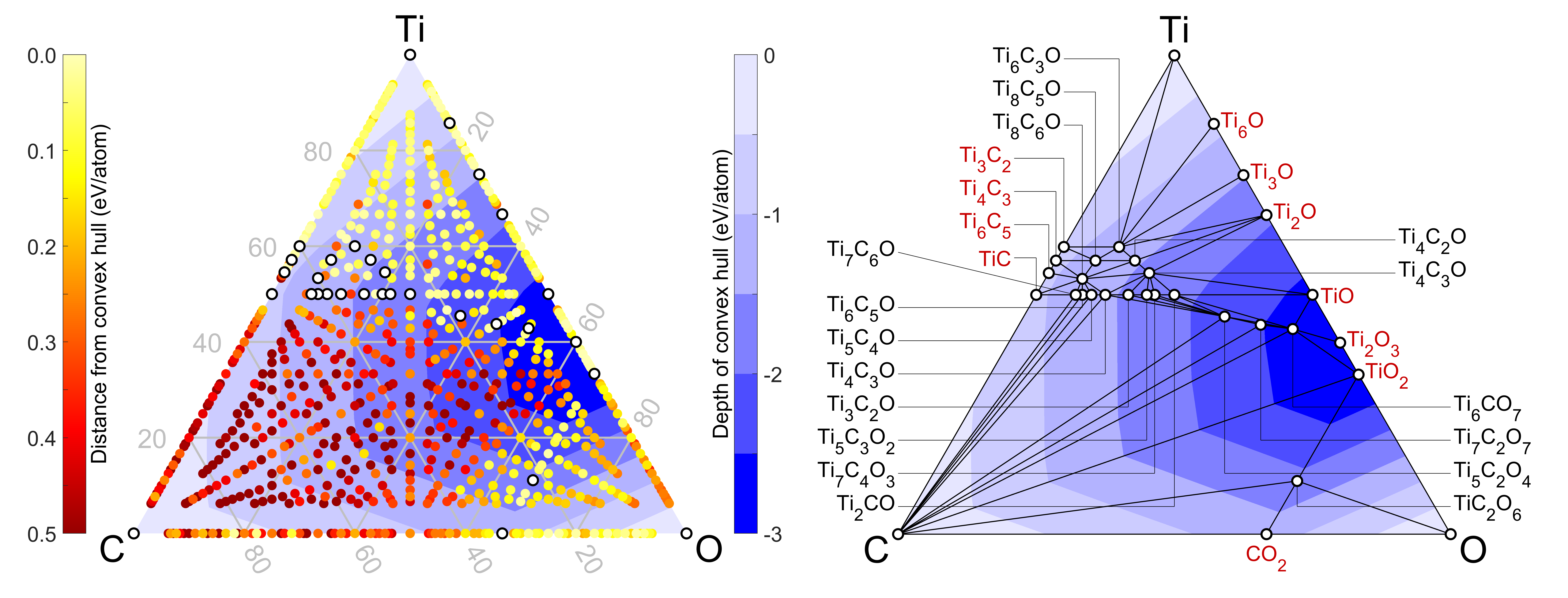}
\caption{\label{fig:Ti_ternary} Calculated Ti-C-O ternary hull using the PBEsol density functional at 0 GPa. (\textit{Left}) Each point corresponds to a particular composition, and is coloured yellow-red based on its distance from the convex hull. For clarity, all compositions that are more than 0.5 eV/atom from the convex hull are set to 0.5 eV/atom on the colour scale. Compositions defining the convex hull are shown in white-filled black circles. (\textit{Right}) The convex hull alone (blue colours), with compositions defining the convex hull labelled by their chemical formulae. Chemical formulae for binary compounds are shown in red, and formulae for elements and ternary compounds are shown in black.}
\end{figure*}

\section{Methods \label{sec:Methods}}        
\subsection{\label{sec:airsssearches} AIRSS searches}  
Crystal structure prediction is carried out using the \textit{ab initio} random structure searching technique (AIRSS) \cite{Pickard-JPCM-2011,AIRSS_2} with version 0.9.1 of the AIRSS code \cite{AIRSS_3}. The AIRSS methodology has been successfully applied in a number of different contexts, such as lithium and sodium-ion battery electrode structures \cite{Mayo_ChemMater_2016,Mayo_ChemMater_2017}, high-pressure binary hydride materials \cite{Fu_ChemMater_2016}, noble-gas compounds \cite{Gao_PRM_2019}, covalent \cite{Sontising_PRM_2020} and ionic solids \cite{Nelson_PRB_2017}.

To illustrate the method, we discuss the \mbox{Ti-C-O} system; considerations for \mbox{Al-C-O} are identical. We explore all compounds \ch{Ti_{$n$}C_{$m$}O_{$p$}} such that $n+m+p\leq 16$. Compounds are taken to be bulk crystalline, with periodic boundary conditions applied in all three dimensions. After removing cases where $n$, $m$ and $p$ are not coprime, this yields a total of 733 unique chemical stoichiometries. The AIRSS approach selects a stoichiometry at random, generates a random structure based on that stoichiometry, and then relaxes its geometry to an enthalpy minimum using (in our case) density-functional theory (DFT). In principle, the chemical species (here Ti, C and O) and the parameter $q$ such that $n+m+p\leq q$, are all that are needed to define the search space. The number of atoms in a generated cell is generally limited to less than 40. Random structures are generated with between 2 and 4 symmetry operations (AIRSS directive \texttt{\#SYMMOPS=2-4}); the precise number is chosen randomly at the moment of generation.

At a bare minimum, we impose distance constraints such that two species are separated by more than the sum of their respective pseudopotential cutoff radii, to prevent poor convergence of the electronic structure. Once a stoichiometry has been sufficiently sampled, we employ automatic pairwise minimum separations, which are derived from the lowest enthalpy structure.

In total, we generate and relax 480,000 \mbox{Ti-C-O} structures and 300,000 \mbox{Al-C-O} structures in our searches. These quantities ensure a thorough sampling of the search space. The underlying structure of first-principles energy landscapes mean that energy basins corresponding to global energy minimum structures tend to be targeted with higher probability by a random search \cite{Pickard-JPCM-2011}. However, AIRSS also affords a quality sampling of metastable phases \cite{Marbella-JACS-2018}, which are important for out-of-equilibrium systems. A significant advantage of the random search approach is that it is trivially parallizable, and can therefore effortlessly harness the parallel power of modern supercomputers, such as the data centers we have used in this study (see Acknowledgements).

To eliminate duplicate structures, we screen for structural similarity using the \texttt{cryan} utility of the AIRSS package, with a similarity threshold of 0.1 \AA. Analysis of space group symmetries is carried out using the tools provided by the \texttt{cellsym}/\texttt{c2x} utility \cite{c2x}, which interfaces with \texttt{spglib} \cite{spglib}.

AIRSS normally proceeds by placing atoms randomly within a unit cell, subject to any symmetry or distance constraints. However, AIRSS can also place entire molecules or molecular fragments randomly instead. These fragments can be derived from techniques which automatically decompose a crystal structure into units \cite{Ahnert-CompMat-2017}. An implementation of the technique in Ref.~\cite{Ahnert-CompMat-2017} is available in the \texttt{cryan} utility of the AIRSS package \cite{AIRSS_3}, and is activated using the \texttt{-g} and \texttt{-bl} flags to \texttt{cryan}. The interested reader can refer to Examples 2.2 and 2.4 of the AIRSS package \cite{AIRSS_3} to see this tool in action. 

We apply this `units-based' approach in our searches. Our workflow sees an inital random search carried out, mapping the convex hull, after which supplemental units-based searches are carried out using the most stable structures appearing on the convex hull. One particularly useful application of this decomposition-based approach is for carbonate structures, where the crystal structure naturally decomposes into planar \ch{[CO3]^{2-}} units and metal ions, and the technique also works well for molecular structures like CO$_2$ and O$_2$. A bond length, set by the \texttt{-bl} flag, is required to define which atoms are considered `connected' when performing the decomposition. For this, we use the minimum bond distance between atoms of any kind in a given structure.

\subsection{\label{sec:ICSD} ICSD crystal structres}   
For comparison and for benchmarking our searches, the Inorganic Crystal Structure Database (ICSD) \cite{ICSD} was queried. At the time of query, the database listed 45 Ti-C binary compounds, 310 Ti-O binary compounds, and 28 C-O binary compounds. In addition, there were 28 entries for elemental Ti, 55 entries for C, and 20 entries for O. There were no listed ternary Ti-C-O compounds. For Al-C-O, the ICSD lists 102 Al-O structures, 5 Al-C structures (and 28 C-O compounds as stated earlier). There are 6 ternary Al-C-O compounds with 4 unique chemical formulae: \ch{Al4O4C}, \ch{Al2CO}, \ch{Al3C15O18}, and \ch{Al6O7C}, and 24 elemental Al entries.

All listed ICSD compounds were downloaded, irregardless of whether these corresponded to atmospheric pressure. Those with fractional occupancies were excluded, and the \texttt{cryan} tool of the AIRSS code was used with a similarity threshold of 0.1 to group identical structures, given that the ICSD can typically contain multiple entries from multiple sources for the same structure. After grouping, we relax the ensuing structures alongside those obtained in our AIRSS searches.

\subsection{Structural relaxations \label{sec:structrelax}}   
Structural relaxations are performed using density-functional theory as implemented in version 18.1 of the CASTEP planewave pseudopotential code \cite{CASTEP}. When carrying out AIRSS searches, we use the PBEsol exchange-correlation functional \cite{PBEsol}, in-built ultrasoft \cite{Vanderbilt-PRB-1990} \texttt{QC5} pseudopotentials for Al, Ti, C and O, a planewave basis set cutoff of 340 eV, and a Monkhorst-Pack \cite{MP_grid} Brillouin-zone sampling density of $<2\pi\times0.07$ \AA$^{-1}$. Relaxations are variable-cell and use CASTEP's implementation of the Broyden-Fletcher-Goldfarb-Shanno (BFGS) algorithm.

The lowest-enthalpy few structures at each stoichiometry are then re-relaxed using the CASTEP code, in-built on-the-fly ultrasoft pseudopotentials \cite{Vanderbilt-PRB-1990}, a planewave basis set cutoff of 850 eV, and a Brillouin-zone sampling density of $<2\pi\times0.03$ \AA$^{-1}$.

Once relaxed, phase stability is calculated for a compound \ch{Ti_{$n$}C_{$m$}O_{$p$}} (or \ch{Al_{$n$}C_{$m$}O_{$p$}}) by computing its formation enthalpy per atom from constituent elements:
\begin{equation}
\label{eq:formation}
\Delta H_f = \frac{H(\mbox{Ti}_n\mbox{C}_m\mbox{O}_p)-nH(\mbox{Ti})-mH(\mbox{C})-pH(\mbox{O})}{n+m+p},
\end{equation}

where $H(x)$ denotes the calculated enthalpy of species $x$. Values of $\Delta H_f$ are then plotted as a function of composition ($n,m,p$), and a convex hull approach yields the stable compositions in the system. 

Eq.~\ref{eq:formation} is a thermodynamic metric for stability, and as such can be sensitive to the level of theory used to compute it. In our case this corresponds to the choice of DFT exchange-correlation functional used in our structural relaxations. Throughout, we use PBEsol, a revised Perdew-Burke-Ernzerhof Generalised Gradient Approximation to the exchange-correlation functional \cite{PBEsol}. A discussion on our choice of functional can be found in Sec.~I of the Supplemental Material and References \cite{SI,Anatase_Howard_1991-s,Brookite_Meagher_1979-s,Columbite_Chen_2002-s,Rutile_Swope_1995-s,CASTEP-DFPT-s,Seekpath-1-s,Seekpath-2-s,Demichelis_CrystEngComm_2012-s,Marques_PCCP_2015-s}. Eq.~\ref{eq:formation} is evaluated at a classical temperature of zero kelvin, and nuclear quantum effects and motion are neglected. Although significant for systems containing light atoms such as hydrogen \cite{Errea-Nature-2016}, we anticipate that nuclear quantum effects will be small for the heavier Ti, Al, C and O atoms considered here. 

\subsection{Data availability}   
Relaxed structures used to generate our convex hulls for the Ti-C-O and Al-C-O systems, input files to \textsc{castep}, and related files, are available on Materials Cloud \cite{materials-cloud-archive}.

\section{Results \label{sec:Results}}          
\subsection{The Ti-C-O ternary system}          

\begin{figure*}
    \centering
    \begin{minipage}{0.666\textwidth}
        \centering
        \includegraphics[width=11.2cm,clip]{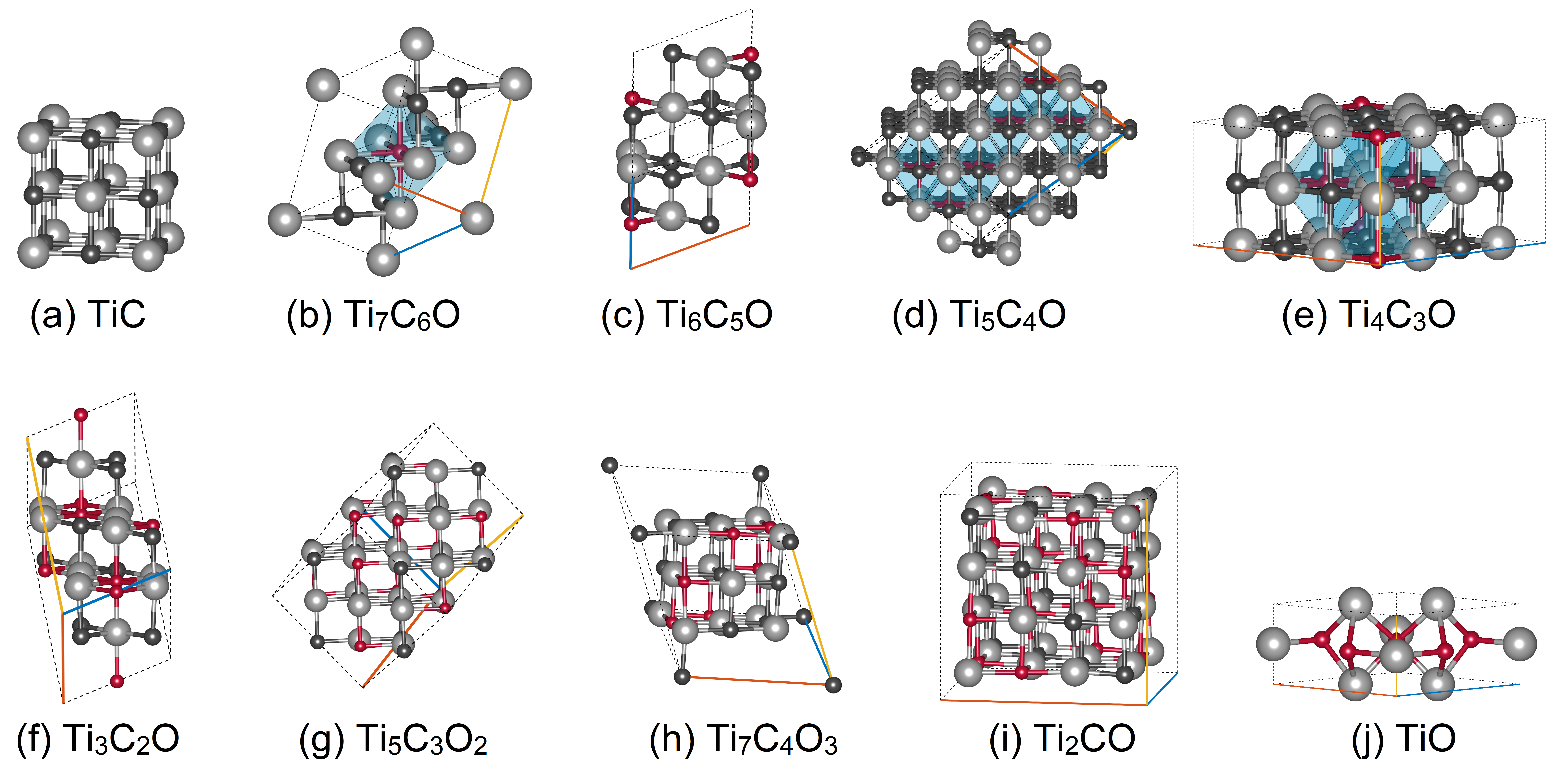}
    \end{minipage}%
    \begin{minipage}{0.333\textwidth}
        \centering
        \includegraphics[width=5.5cm,clip]{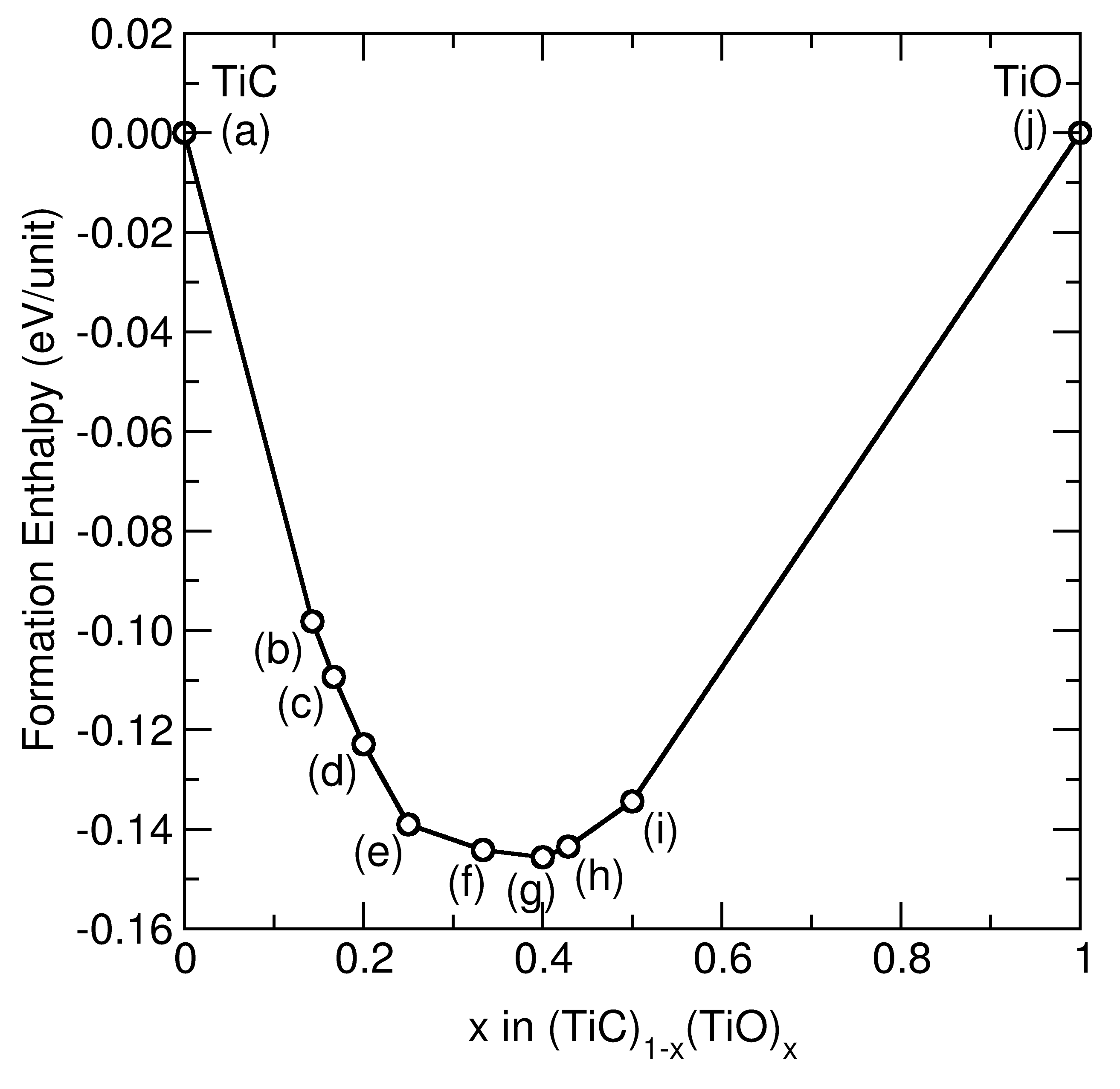}
    \end{minipage}
\caption{\label{fig:ae_belt} (\textit{Left}) Predicted crystal structures of Ti-C-O compounds in the `oxycarbide belt' region of the convex hull, \textit{i.e.} the \ch{TiC}$-$\ch{TiO} pseudobinary (see Fig.~\ref{fig:Ti_ternary}). Titanium atoms are shown in light grey, carbon atoms in dark grey, and oxygen atoms in red. Blue polyhedra where drawn are oxygen coordination polyhedra. (a) Titanium carbide, \ch{TiC}, space group $Fm\overline{3}m$, with the rocksalt structure. (b) \ch{Ti7C6O}, with space group $R\overline{3}$. (c) \ch{Ti6C5O}, with space group $P\overline{1}$. (d) \ch{Ti5C4O}, with space group $P\overline{1}$. (e) \ch{Ti4C3O}, with space group $P4_2 / mnm$. (f) \ch{Ti3C2O}, with space group $C2/c$. (g) \ch{Ti5C3O2}, with space group $R\overline{3}m$. (h) \ch{Ti7C4O3}, with space group $R\overline{3}$. (i) \ch{Ti2CO}, with space group $Fd\overline{3}m$. (j) \ch{TiO}, with space group $P\overline{6}2m$. (\textit{Right}) \ch{TiC}$-$\ch{TiO} pseudobinary convex hull showing the formation enthalpy of compounds (a)$-$(j) from \ch{TiC} and \ch{TiO}.}
\end{figure*}

The results of our searches on the \mbox{Ti-C-O} system are shown in Fig.~\ref{fig:Ti_ternary}. Coloured circles in that figure represent the various stoichiometries searched, one for each stoichiometry. Compounds which define the convex hull are shown as black-ringed white circles. 

As far as elemental Ti, C and O are concerned, the hcp and graphite structures for Ti and C are straightforwardly reproduced in our searches. Experimentally, at low temperatures, solid oxygen forms molecular, magnetic insulating phases such as the $\alpha$ (space group $C2/m$) and structurally related $\delta$ phases (space group $Fmmm$) \cite{Klotz-PRL-2010}. However, DFT calculations, including those of the present work, favour nonmagnetic metallic structures such as $Cmcm$ oxygen \cite{Neaton-PRL-2002,Nelson_PCCP_2015}, which feature either clusters or linear chains of oxygen molecules, and which are more in line with the higher pressure $\epsilon$ oxygen phase observed in experiments \cite{Lundegaard-Nature-2006,Fujihisa-PRL-2006}. We proceed with the lowest enthalpy oxygen structure found in our searches when using Eq.~\ref{eq:formation}. 

In terms of the binary titanium oxides, we find in order of increasing Ti content that \ch{TiO2}, \ch{Ti2O3}, \ch{TiO}, \ch{Ti2O}, \ch{Ti3O} and \ch{Ti6O} are all stable. The anatase polymorph of \ch{TiO2} (space group $I4_1/amd$) is predicted to be energetically most favourable; this is also known to be the case when using accurate Quantum Monte Carlo methods \cite{Trail_PRB_2017}. All six of these titanium oxides are known in the ICSD.

We find four stable titanium carbides: \ch{TiC}, \ch{Ti6C5}, \ch{Ti4C3} and \ch{Ti3C2} (in order of increasing Ti content). \ch{TiC} is predicted to take on the rocksalt $Fm\overline{3}m$ structure, as indeed observed experimentally. The remaining stoichiometries \ch{Ti6C5}, \ch{Ti4C3} and \ch{Ti3C2} have been reported in previous structure prediction studies \cite{Xu_RScAdv_2018}. Carbon dioxide, \ch{CO2}, is the only predicted stable oxide of carbon, and is found in our searches in its experimentally observed, $Pa3$-symmetry `Phase I' structure \cite{Bonev_PRL_2003}. It is again worth emphasising again that the known or previously discovered phases and compositions discussed here are reproduced by our approach without recourse to prior crystal structures or databases.

Moving away from binary compounds, we identify 17 stable ternary compounds in Fig.~\ref{fig:Ti_ternary}. The red-yellow colour coding in the left panel of Fig.~\ref{fig:Ti_ternary} reveals areas of interest. The upper triangular region defined by titanium contents exceeding 50\% near the top of the left panel of Fig.~\ref{fig:Ti_ternary}, and the region in the lower-right portion of that same panel, defined by oxygen contents exceeding about 60\%, collectively contain all our predicted stable ternary phases. In these regions, structures typically lie within \mbox{0.0$-$0.2 eV/atom} of the convex hull. On the other hand, the lower-left carbon-rich part of the convex hull diagram tends to consist of unstable compositions far from the convex hull; the convex hull itself (blue shading in Fig.~\ref{fig:Ti_ternary}) is also shallow in this region. 

The `oxycarbide' belt region, i.e.~the \ch{TiC}$-$\ch{TiO} pseudobinary, contains eight stable ternary structures, excluding the \ch{TiC} and \ch{TiO} endmembers. These structures are shown in Fig.~\ref{fig:ae_belt}(a)-(j), with the \ch{TiC}$-$\ch{TiO} pseudobinary convex hull shown in the right-hand panel of Fig.~\ref{fig:ae_belt}. Apart from \ch{TiO}, whose lowest-enthalpy (stoichiometric) structure is hexagonal in our calculations (`$\epsilon$-TiO' \cite{Amano_ACIE_2016}), all other compounds on the oxycarbide belt are pseudocubic and are based on the rocksalt \ch{TiC} structure, with increasing numbers of C atoms substituted for O. This is in accordance with the description of these phases in the experimental work of Miller \textit{et al.}~\cite{Miller_JMCA_2016}. In practice, experimentally, these compounds are actually entropically-stabilized solid solutions of the form TiC$_{1-x}$O$_x$, and what AIRSS finds are periodically-ordered approximants to those solid solutions (Fig.~\ref{fig:ae_belt}(a)-(j)). As can be seen in Fig.~\ref{fig:ae_belt}, we do not find any structures on the convex hull for $x>0.5$, however we do find metastable phases which are also pseudocubic.

\subsection{Titanium carbonate and carbonate formation}          

\begin{figure}
\centering
\includegraphics[width=8.0cm]{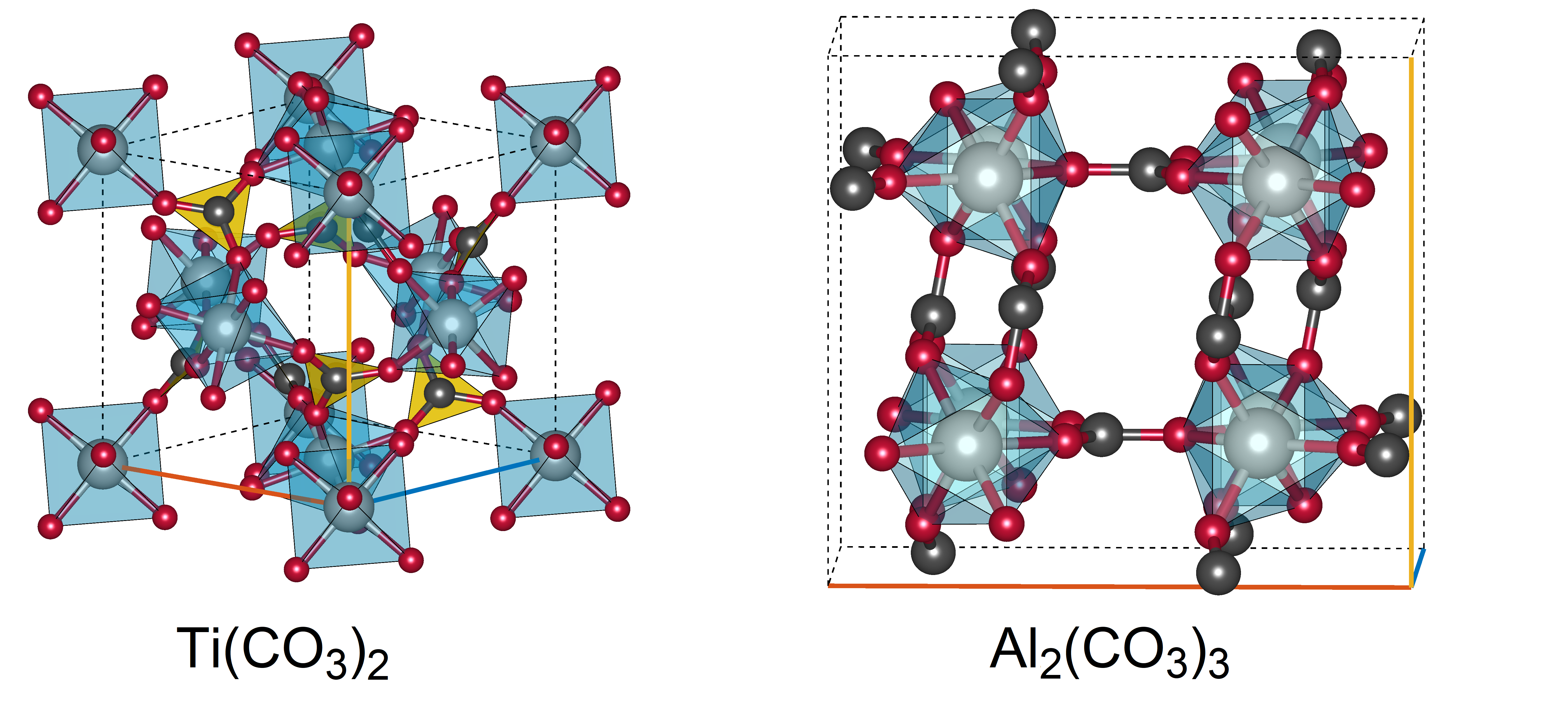}
\caption{\label{fig:carbonates} (\textit{Left}) Predicted structure and unit cell of titanium carbonate, \ch{Ti(CO3)2}, with space group $Pa3$. Carbonate \ch{[CO3]^{2-}} units are shown as yellow triangles, with oxygen atoms in red and carbon atoms in dark grey. Titanium atoms are shown in light grey. \ch{TiO6} octahedra (light blue) occupy the corners and face centers of the cubic unit cell, and are linked by C atoms. (\textit{Right}) Predicted structure and unit cell of aluminium carbonate, \ch{Al2(CO3)3}, with space group $Cmcm$. Aluminium atoms are in light grey, carbon atoms are in dark grey, and oxygen atoms are in red.}
\end{figure}

In the oxygen-rich part of the phase diagram, we predict titanium carbonate, \ch{Ti(CO3)2} (\ch{TiC2O6}), to be stable. The predicted structure of titanium carbonate is cubic, space group $Pa3$, and its structure is depicted in Fig.~\ref{fig:carbonates}. While there are several known examples of transition metal carbonates, such as \ch{MnCO3}, \ch{FeCO3} and \ch{CoCO3}, titanium carbonate has not to our knowledge been discussed before. A 19th century paper does however detail a `carbonate of titanium' medicinal remedy, but this turned out to be largely ferrous sulfate and ferric oxide \cite{JAMA_1884}. The calculated phonon spectrum of \ch{Ti(CO3)2} is provided in Sec.~II of the Supplemental Material and References \cite{SI,Anatase_Howard_1991-s,Brookite_Meagher_1979-s,Columbite_Chen_2002-s,Rutile_Swope_1995-s,CASTEP-DFPT-s,Seekpath-1-s,Seekpath-2-s,Demichelis_CrystEngComm_2012-s,Marques_PCCP_2015-s}; a lack of imaginary phonon frequencies confirms the dynamic stability of this predicted compound. Concerning technological applications, transition metal carbonates (\ch{MCO3}) have been suggested as a replacement for transition metal oxides (\ch{MO}) as lithium ion battery components, due to their higher prospective capacities for lithium \cite{Wang_AMI_2014}. The existence of titanium carbonate is chemically intuitive in the sense that it satisfies the octet rule as \ch{Ti^{4+}(CO3^{2-})2}, suggesting that titanium is in the +4 (IV) oxidation state. We also remark that the stability of titanium carbonate hints at the stability of other `missing' Group 4 (IVb) carbonates such as hafnium and zirconium carbonate, and preliminary work of ours indeed supports this idea. Structure prediction has played an important role in the elucidation of carbonate structures, for example in MgCO$_3$ at high pressures \cite{Binck_PRM_2020}.

Carbonate compounds are also of interest in processes which attempt to capture \ch{CO2} through carbon mineralisation, usually involving calcium or magnesium carbonates. For \ch{CaCO3} and \ch{MgCO3}, the reaction of the metal oxide (\ch{CaO} or \ch{MgO}) with \ch{CO2} to form the carbonate is strongly exothermic, and is usually only reversible at high temperatures. Quantatively, for the reaction \ch{CaO + CO2 -> CaCO3}, we calculate \mbox{$\Delta H = -157$ kJ/mol} (aragonite) and \mbox{$\Delta H = -160$ kJ/mol} (calcite), which may be compared to the measured value of \mbox{$\Delta H = -179$ kJ/mol} \cite{Lackner_Energy_1995}, while for \ch{MgO + CO2 -> MgCO3} we calculate \mbox{$\Delta H = -106$ kJ/mol} compared to a measured value of \mbox{$\Delta H = -118$ kJ/mol}. For \ch{Ti(CO3)2}, we instead find a much smaller value: \mbox{$\Delta H = -3$ kJ/mol} for \ch{1/2 TiO2 + CO2 -> 1/2 Ti(CO3)2}, so that the reaction of the metal oxide with \ch{CO2} is much closer to equilibrium than it is for \ch{CaCO3} and \ch{MgCO3}, suggesting a use for \ch{Ti(CO3)2} as a \textit{reversible} carbon capture material.

The calculations in the preceding paragraph do not consider the effects of temperature. We can, at least partially, take temperature into account using the quasiharmonic approximation. We find that temperature tends to inhibit the formation of titanium carbonate in the quasiharmonic approximation. For the reaction \ch{1/2 TiO2 + CO2 -> 1/2 Ti(CO3)2}, we calculate \mbox{$\Delta G = +3$ kJ/mol} at $T = 0$ K (taking into account zero-point energy), and \mbox{$\Delta G = +13$ kJ/mol} at $T = 300$ K. The forward reaction is therefore likely to be favoured at \textit{low} temperature and (as we discuss next) \textit{high} pressure.

It is worth exploring the effect of pressure on these reactions as well. The change in reaction enthalpy with respect to pressure for the formation of a carbonate (\ch{MCO3}) from a corresponding metal oxide (\ch{MO}) and \ch{CO2} at 0 GPa, namely $d(\Delta H)/dP|_{\mbox{\footnotesize{0 GPa}}}$, is expressible in terms of the volume of formation \cite{Pickard-JPCM-2011}: $d(\Delta H)/dP |_{\mbox{\footnotesize{0 GPa}}} = \Delta V$, with $\Delta V = V_{\mbox{\footnotesize{MCO3}}}-V_{\mbox{\footnotesize{MO}}}-V_{\mbox{\footnotesize{CO}}_2}$. Typically, the carbonate \ch{MCO3} is significantly denser, per mol, than its constituent oxide (\ch{MO}) and \ch{CO2} components, and so the application of pressure decreases $\Delta H$ and favours the formation of the carbonate.

\begin{figure}
\centering
\includegraphics[width=8.0cm,clip]{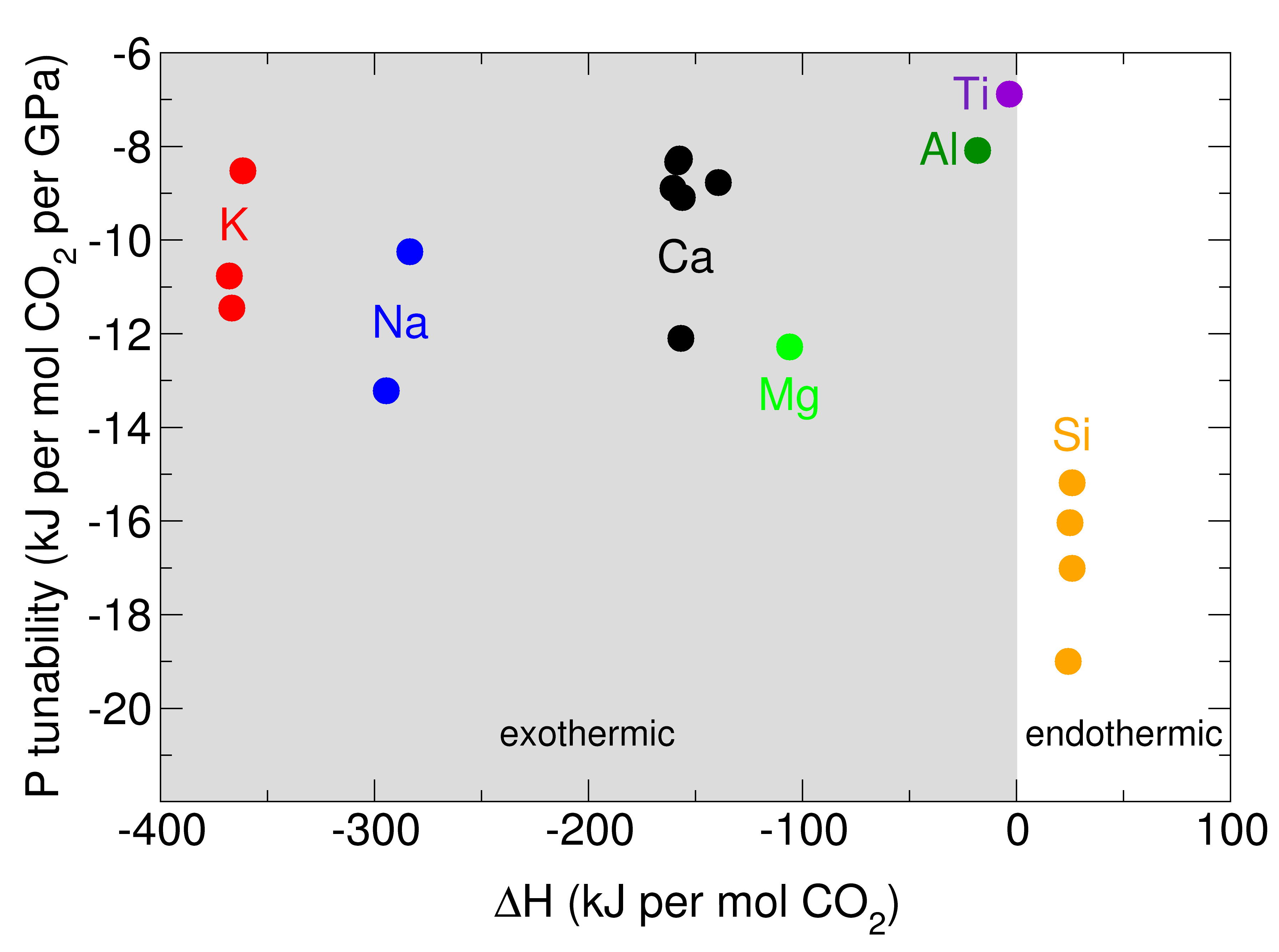}
\caption{\label{fig:othercarbonates} Plot showing the reaction enthalpies, $\Delta H$, for the formation of various carbonate compounds from their respective oxides and \ch{CO2} compared with pressure tunabilities at 0 GPa, $d(\Delta H)/dP|_{\mbox{\footnotesize{0 GPa}}}$. For the crystal structures of oxides and carbonates appearing in this plot, see the Supplemental Material and References \cite{SI,Anatase_Howard_1991-s,Brookite_Meagher_1979-s,Columbite_Chen_2002-s,Rutile_Swope_1995-s,CASTEP-DFPT-s,Seekpath-1-s,Seekpath-2-s,Demichelis_CrystEngComm_2012-s,Marques_PCCP_2015-s}.}
\end{figure}

Given a metal M, we therefore have two metrics: the change in enthalpy $\Delta H$, per mol of \ch{CO2}, for the reaction \ch{MO + CO2 -> MCO3}, and the extent to which $\Delta H$ can be changed by pressure, namely $d(\Delta H)/dP|_{\mbox{\footnotesize{0 GPa}}}$. We refer to the latter quantity as the \textit{pressure tunability}. In Fig.~\ref{fig:othercarbonates}, we plot these quantities for a variety of carbonates, namely M = K, Na, Ca, Mg, Ti, Si and Al; we will also discuss the case \mbox{M = Al} in more detail shortly. The crystal structures used for these metal oxides and carbonates are given in Sec.~III of the Supplemental Material and References \cite{SI,Anatase_Howard_1991-s,Brookite_Meagher_1979-s,Columbite_Chen_2002-s,Rutile_Swope_1995-s,CASTEP-DFPT-s,Seekpath-1-s,Seekpath-2-s,Demichelis_CrystEngComm_2012-s,Marques_PCCP_2015-s}. The pressure tunability is quite similar for all carbonates considered, however, the compounds \mbox{M = Ti, Al} lie closest to the line $\Delta H = 0$, signifying the fact that $\Delta H$ can be tuned, and changed from positive to negative, for the reaction \ch{MO + CO2 -> MCO3} using a modest amount of pressure. Overall, this suggests at a possible `pressure cycle' for capturing \ch{CO2}: applying pressure to \ch{TiO2} favours the reaction with \ch{CO2} and carbonate formation, while releasing pressure may allow recovery of the original \ch{TiO2} material and the release of \ch{CO2}.

\begin{figure*}
\centering
\includegraphics[width=17.0cm]{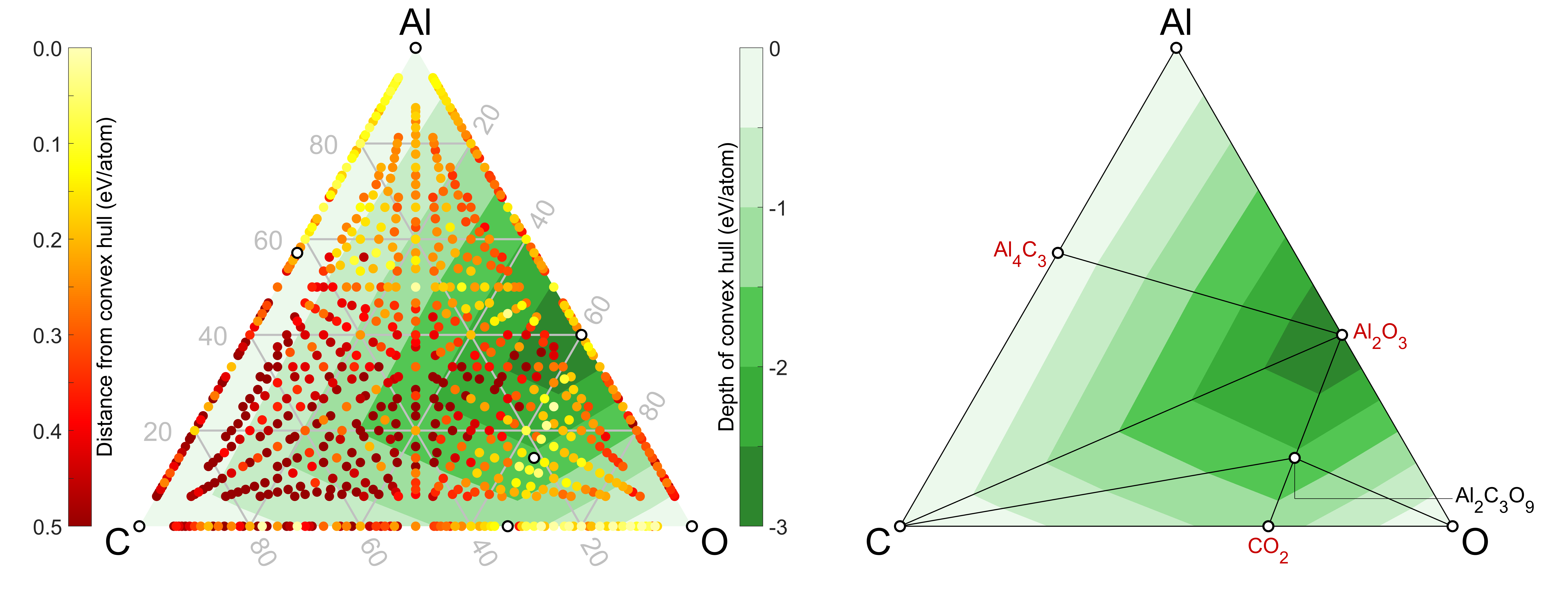}
\caption{\label{fig:Al_ternary} Calculated Al-C-O ternary hull using the PBEsol density functional at 0 GPa. (\textit{Left}) Each point corresponds to a particular composition, and is coloured yellow-red based on its distance from the convex hull. For clarity, all compositions that are more than 0.5 eV/atom from the convex hull are set to 0.5 eV/atom on the colour scale. Compositions defining the convex hull are shown in white-filled black circles. (\textit{Right}) The convex hull alone (green colours), with compositions defining the convex hull labelled by their chemical formulae. Chemical formulae for binary compounds are shown in red, and formulae for elements and ternary compounds are shown in black.}
\end{figure*}

Finally, we discuss nanostructuring, in particular porous alkali metal carbonates such as mesoporous magnesium carbonate (MMC). MMC can be synthesised and shows good moisture and gas sorption, on par with advanced zeolites \cite{Forsgren-PLOSONE-2013}. Should titanium carbonate be able to be synthesized as a mesoporous carbonate, this offers an additional application for this material.

\subsection{The Al-C-O ternary system \label{sec:Al-C-O}}          
Moving our discussion to the Al-C-O system, the results of our searches on the \mbox{Al-C-O} system are shown in Fig.~\ref{fig:Al_ternary}. 

For the elements, the fcc structure of Al is reproduced; the structures for carbon and oxygen are as discussed earlier. The only predicted stable oxide is \ch{Al2O3}, in the $R\overline{3}c$-symmetry corundum structure. The formation of oxides \ch{Al2O} and \ch{AlO} has been reported experimentally at high temperatures ($>$1000 $^{\circ}$C) \cite{Hoch_JACS_1954} and the corresponding structures are listed in the ICSD, however, these compounds disproportionate into Al and \ch{Al2O3} upon cooling. The only stable aluminium carbide is \ch{Al4C3}, whose structure is reported as rhombohedral \cite{Gesing_ZB_1995}. Our searches recover that experimental structure, as well as slightly more stable ($\approx$2.7 kJ/mol) cubic and orthorhombic modifications of \ch{Al4C3}.

Moving out into ternary compounds, the situation is quite different from the Ti-C-O system. Our calculations show a single stable phase: aluminium carbonate, \ch{Al2(CO3)3} (Fig.~\ref{fig:Al_ternary}). We are not aware of any definitive characterisation of aluminium carbonate in the literature, however, carbonate species are known to form at the surfaces of the $\gamma$ modification of \ch{Al2O3} \cite{Szanyi_PCCP_2014}. As was the case for titanium carbonate, this compound fulfils the octet rule, this time with Al in a +3 oxidation state. The predicted structure of aluminium carbonate is orthorhombic, space group $Cmcm$, and is depicted in Fig.~\ref{fig:carbonates}. The calculated phonon spectrum of \ch{Al2(CO3)3} is provided in Sec.~II of the Supplemental Material and References \cite{SI,Anatase_Howard_1991-s,Brookite_Meagher_1979-s,Columbite_Chen_2002-s,Rutile_Swope_1995-s,CASTEP-DFPT-s,Seekpath-1-s,Seekpath-2-s,Demichelis_CrystEngComm_2012-s,Marques_PCCP_2015-s}; as was the case for \ch{Ti(CO3)2}, a lack of imaginary phonon frequencies confirms the dynamic stability of this predicted compound. A point of difference between titanium and aluminium carbonate is that the former contains no Ti$-$O$-$Ti bonds: the \ch{TiO6} octahedra in its structure (Fig.~\ref{fig:carbonates}) do not share corners. On the other hand, the predicted structure of aluminium carbonate does contain Al$-$O$-$Al bonds.

For the reaction \ch{1/3 Al2O3 + CO2 -> 1/3 Al2(CO3)3}, we calculate \mbox{$\Delta H = -18$ kJ/mol} at the static-lattice level. The reaction remains favourable even when temperature is taken into account through the quasiharmonic approximation, although it is inhibited by temperature: we find \mbox{$\Delta G = -10$ kJ/mol} at \mbox{$T = 0$ K} (including zero-point energy), and \mbox{$\Delta G = -6$ kJ/mol} at \mbox{$T = 150$ K}, but \mbox{$\Delta G = +1$ kJ/mol} at \mbox{$T = 300$ K}. As was the case for titanium carbonate, we therefore conclude that the forward reaction is likely to be favoured by low temperatures and high pressures. 

Our search results can be straightforwardly queried to see if other reactions between metal oxides and carbon dioxide are possible. In the case of the Ti-C-O system, the only compound we calculate to form exothermically from \ch{TiO2} and \ch{CO2} is titanium carbonate, \ch{Ti(CO3)2}. However, in the Al-C-O system, there is more than one compound which could form from \ch{Al2O3} and \ch{CO2}: both the carbonate, \ch{Al2(CO3)3}, and \ch{Al2CO5}, could form. Said differently, both of these compounds are calculated to be stable with respect to decomposition into \ch{Al2O3} and \ch{CO2}. Of the two, only \ch{Al2(CO3)3} sits on the \ch{Al2O3}$-$\ch{CO2} pseudobinary convex hull, so \ch{Al2CO5} would be considered metastable. 

Unlike the \mbox{Ti-C-O} system, the ICSD lists some ternary \mbox{Al-C-O} compounds. All of these lie on the \ch{Al2O3}$-$\ch{Al4C3} pseudobinary, i.e., are of the form \ch{Al_{4-2$x$}O_{3$x$}C_{3-3$x$}} for $x=1/2$ (\ch{Al2CO}), $x=4/5$ (\ch{Al4O4C}), and $x=7/8$ (\ch{Al6O7C}). There is an exception - the ICSD lists a MOF structure with chemical formula \ch{Al3C15O18} \cite{Reinsch_CAEJ_2017}. This MOF has actual composition \ch{[Al(OH)(O2C-C3H4-CO2)] * $n$ * H2O}, and the ICSD structure does not contain H positions. With $n=1$ and excluding hydrogen, this is \ch{AlO5C6} (i.e. \ch{Al3C15O18}). \ch{Al3O_{3.5}C_{0.5}} (or \ch{Al6O7C}) has been synthesised in Ref.~\citenum{Asaka_IC_2013}; it is described as an intergrowth structure with space group $R\overline{3}m$, consisting of single layers of $\alpha$-\ch{Al2O3} and \ch{Al2(OC)4} layers. Its listed structure has fractional occupancies. Three structures, all with fractional occupancies, have been suggested for aluminium monoxycarbide \ch{Al2CO}: $\alpha-$, $\alpha'-$ and $\alpha''$ \cite{Grass_JSC_2004}. In our case, AIRSS finds a structure 18 meV/atom above the convex hull, with $Pca2_1$ symmetry.

\section{Discussion \label{sec:Discussion}}      
The technologically-relevant Ti-C-O and Al-C-O ternary systems support a rich variety of stable compounds. We have shed new light on these systems thorugh computational means, using extensive random structure searching combined with a `units-based' approach in which structures are built from molecular fragments. Our approach to these systems has no adjustable parameters, is straightforward to set up, and requires no prior knowledge. In carrying out this work, one of our goals is that the techniques detailed here will inspire the use of structure prediction calculations to push the frontiers of materials science out into ternary systems and beyond.

Known compounds in the Ti-C-O and Al-C-O systems, such as familiar titanium or aluminum oxides and carbides, are readily rediscovered in this work, in their correct, experimentally-known structures, supporting the viability of our approach.

Full ternary searches detect 17 stable ternary compounds (i.e., containing Ti, C and O) in the Ti-C-O system (Fig.~\ref{fig:Ti_ternary}), and 1 stable ternary compound in the Al-C-O system. Of these, 8 lie along the oxycarbide `belt' region of the Ti-C-O system, which is to say on the \ch{TiC}$-$\ch{TiO} pseudobinary line (Fig.~\ref{fig:ae_belt}). Our ternary convex hull underscores the importance of carrying out a full ternary search, as we find stable compositions which do not lie along pseudobinary lines.

An interesting discovery is that both the Ti-C-O and Al-C-O systems support stable carbonates: \ch{Ti(CO3)2} and \ch{Al2(CO3)3} (Fig.~\ref{fig:carbonates}). These compounds fulfil the octet rule and as such, their existence could have been argued based on chemical intuition. In our case, their stability falls out naturally as a consequence of our systematic approach. Carbonates provide excellent test-cases for our `units-based' methodology; the reason for this is that our searches detect that both \ch{Ti(CO3)2} and \ch{Al2(CO3)3} contain planar carbonate \ch{[CO3]^{2-}} units, allowing us to carry out additional searches by building new structures with either Ti or Al atoms in conjunction with carbonate ion building blocks.

Carbonate minerals can sequester carbon dioxide through carbon mineralisation. Calcium and magnesium carbonates, being comparably abundant in Earth's crust and mantle, have been widely studied for this purpose. For both of these minerals, the reaction \ch{MO + CO2 -> MCO3} is strongly exothermic, but can be reversed at high temperatures. For titanium and aluminium carbonate however, we calculate that this reaction is close to equilibrium (i.e. $\Delta H \approx 0$), suggesting that the reaction could be controlled and reversed quite easily using temperature and pressure. In this way, we envisage titanium and aluminum carbonates being used as carbon capture materials, where the original material is recoverable and therefore reusable. The application of high pressure is calculated to favour the forward reaction (formation of the carbonate), a feature common among most carbonate minerals (Fig.~\ref{fig:othercarbonates}).

When examining the calculated ternary convex hulls for the Ti-C-O and Al-C-O systems (Figs.~\ref{fig:Ti_ternary} and \ref{fig:Al_ternary}), it is useful to keep several things in mind, particularly when comparing results with other studies or with experimental work. To illustrate, let us discuss and compare our Al-C-O phase diagram (Fig.~\ref{fig:Al_ternary}) with the Al-C-O phase diagram generated by the Materials Project \cite{MaterialsProject}. At the time of writing, the phase diagram app \cite{MPpda1,MPpda2} of the Materials Project produces a phase diagram with 3 stable binary phases (\ch{CO2}, \ch{Al4C3} and \ch{Al2O3}), and 1 stable ternary phase: \ch{Al4CO4}. Our phase diagram (Fig.~\ref{fig:Al_ternary}) agrees with this in terms of stable binary phases, however in our phase diagram, \ch{Al4CO4} is metastable and lies 33 meV/atom above the convex hull, and the only stable ternary phase is \ch{Al2(CO3)3}. There are several things to consider here. The first is coverage: our phase diagram considers 733 unique stoichiometries (Sec.~\ref{sec:airsssearches}), whereas the Materials Project has 26 unique stoichiometries in the Al-C-O system, and does not have a \ch{Al2(CO3)3} structure with which to make a comparison. The second is level of theory: we have used the PBEsol density functional (Sec.~\ref{sec:structrelax}), whereas total energies in the Materials Project are evaluated with the PBE functional \cite{PBE}. Third, the Materials Project applies an energy correction to certain classes of materials, such as oxides and superoxides, with the goal of producing more accurate formation energies \cite{Wang_chemarxiv_2021}; in this work, such corrections aren't applied and all compounds are treated at the same level of theory. Finally, the most stable crystal structure for a given compound can differ between the present work and the Materials Project. To illustrate, the most stable structure for \ch{Al4C3} reported in the Materials Project is the experimentally-known rhombohedral structure, whereas in the present work, AIRSS found cubic and orthorhombic modifications of \ch{Al4C3} which were calculated to be more stable (see the discussion in Sec.~\ref{sec:Al-C-O}).

In light of these factors, it should be pointed out that it is not only the stable phases in Figs.~\ref{fig:Ti_ternary} and \ref{fig:Al_ternary} (i.e., the ones defining the convex hull) which are important, but also the low-lying metastable phases as well. Such phases may certainly be synthesizable, even if the calculations of the present work do not place those compounds exactly on the convex hull. A good example would be familiar carbon monoxide (CO), which does not sit on the convex hull (Figs.~\ref{fig:Ti_ternary} and \ref{fig:Al_ternary}).

Finally, it is worth briefly revisiting and discussing the ICSD structures which we considered alongside our searches (Sec.~\ref{sec:ICSD}). Note that the ICSD can contain multiple entries for the same crystal structure. Of the ICSD structures considered in the Ti-C-O system, and which relax successfully in our calculations, we end up with 45 unique binary structures (and 0 ternary structures). Of these 45 structures, 5 correspond to binaries which lie outside the scope of our search as defined in Sec.~\ref{sec:airsssearches} (e.g.~Ti$_9$O$_{17}$), leaving 40 structures which are potentially discoverable using AIRSS. 

In 19 cases, AIRSS found the ICSD crystal structure as-is, and in 20 cases, AIRSS did not find the ICSD structure, but did find a more stable structure at that stoichiometry. Note that of the latter 20 cases, 9 were high-energy metastable structures ($>$ 0.2 eV/f.u., and in some cases $>$ 1.0 eV/f.u. above the ground state), and so are unlikely targets for AIRSS, which is biased towards low-energy structures with large energy basins of attraction. There was only 1 structure which AIRSS did not find and which AIRSS did not also find a more stable structure. There were no ICSD structures on the convex hull which AIRSS did not also find.

In the Al-C-O system, we end up with 8 unique binary ICSD structures, and 1 ternary structure. Of these, AIRSS found 4 of the structures as-is (including the ternary structure), and in the remaining 5 cases, AIRSS did not find the structure, but did find a more stable structure at the same stoichiometry. In these latter 5 cases, 4 were high-energy metastable phases.

\section{Conclusions \label{sec:Conclusions}}      
The Ti-C-O and Al-C-O ternary systems have been mapped out using extensive first-principles structure prediction calculations, in which more than 780,000 structures were relaxed. Familiar oxides and carbides are recovered in our approach, emerging naturally from our structure searches without recourse to prior knowledge. These systems also play host to many ternary compounds; in particular, our calculations reveal the stability of octet-rule-fulfiling titanium and aluminium carbonates, \ch{Ti(CO3)2} and \ch{Al2(CO3)3}, respectively. In this way, we have `computationally synthesized' these compounds through purely calculational means. An examination of the formation enthalpies for these carbonates from their respective metal oxides demonstrates that these materials may be able to act as carbon sequesting compounds.

\section{Acknowledgements}  
All authors are grateful for computational support from the UK national high performance computing service, ARCHER, and for computational resources from the UK Materials and Molecular Modelling Hub. Access to these facilities was provided through the UKCP consortium, EPSRC grants EP/P022561/1, EP/P020194/1, and EP/T022213/1. JRN acknowledges financial support from the Joint Research Center and the Advanced Target Project at AIMR, Tohoku University. CJP acknowledges financial support from EPSRC grant EP/P022596/1, and a Royal Society Wolfson research merit award. RJN acknowledges financial support from EPSRC grant EP/J017639/1.

\section{References}     


\pagebreak

\widetext

\newpage

\large 

\centerline{\textbf{Supplemental Material for:}}

\vspace{0.1cm}

\centerline{\textbf{``Navigating the Ti-C-O and Al-C-O ternary systems through theory-driven discovery''}}

\vspace{0.1cm}

\centerline{\textbf{J.~R.~Nelson,$^{1,2*}$ R.~J.~Needs,$^3$ and C.~J.~Pickard$^{1,2}$}}

\vspace{0.1cm}

\normalsize
\centerline{\textit{$^1$Department of Materials Science and Metallurgy, University of
Cambridge,}}

\centerline{\textit{27 Charles Babbage Road, Cambridge CB3 0FS, United Kingdom}}

\centerline{\textit{$^2$Advanced Institute for Materials Research, Tohoku University, }}

\centerline{\textit{2-1-1 Katahira, Aoba, Sendai, 980-8577, Japan}}

\centerline{\textit{$^3$Theory of Condensed Matter Group, Cavendish Laboratory,}}

\centerline{\textit{J.~J.~Thomson Avenue, Cambridge CB3 0HE, United Kingdom}}

\vspace{0.1cm}

\centerline{$^{*}$Email: \texttt{jn336@cam.ac.uk}}


\setcounter{section}{0}
\setcounter{page}{1}
\makeatletter

\section{On the use of the PBEsol exchange-correlation functional} 
A choice must be made as to the level of theory which balances accuracy and computational cost. For this work we used the Perdew-Burke-Ernzerhof for solids `PBEsol' functional which is designed to improve the equilibrium properties of densely packed solids and their surfaces \cite{PBEsol-x}.

\begin{figure}[h]
\centering
\includegraphics[width=12.0cm]{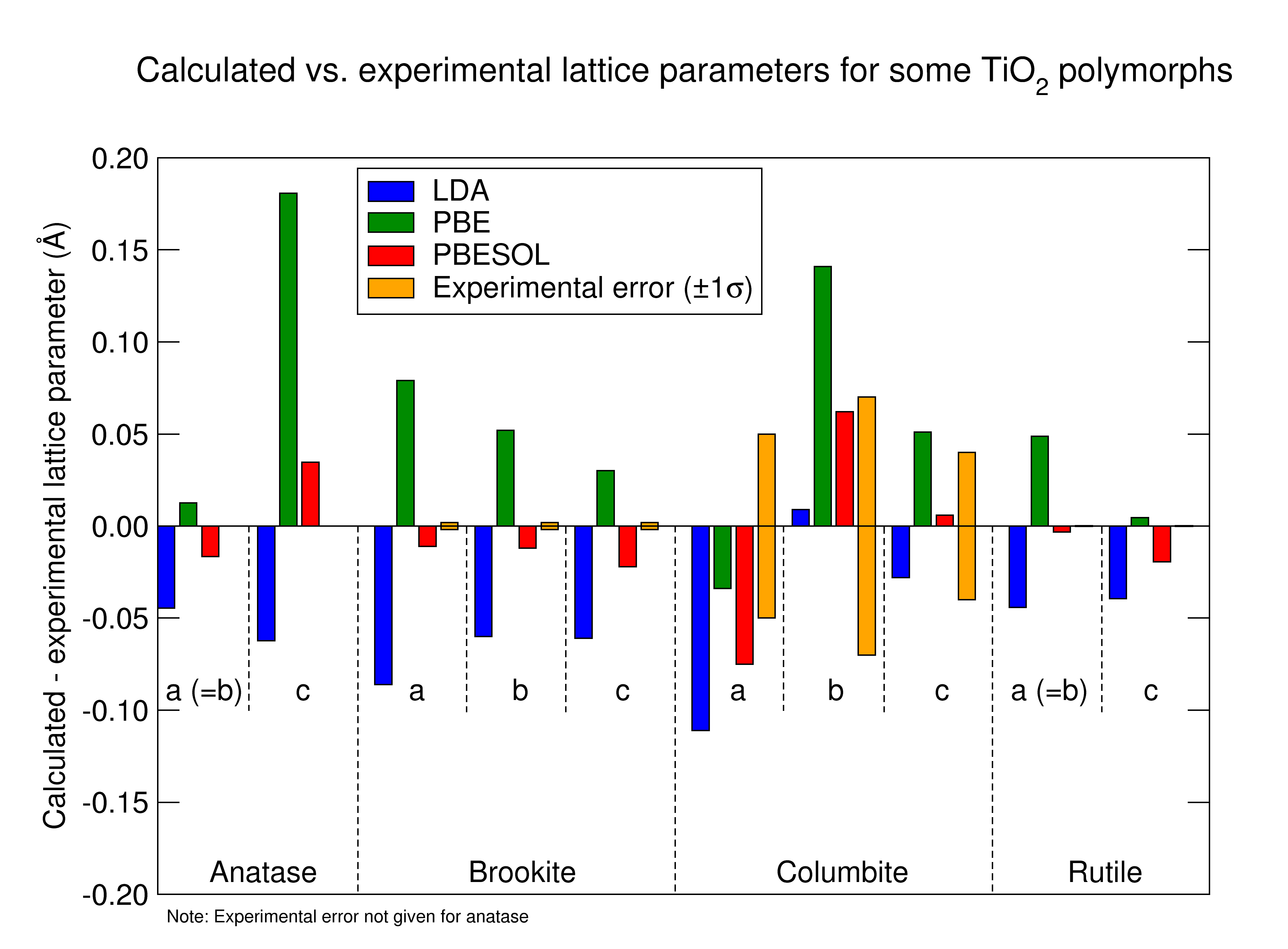}
\caption{\label{fig:Latt_params_TiO2} Lattice parameters for bulk TiO$_2$ structures as calculated using three different xc-functionals (LDA, PBE, PBEsol) compared to experimental measurements of those lattice parameters. Experimental values are taken from Refs.~\cite{Anatase_Howard_1991-x} (anatase), \cite{Brookite_Meagher_1979-x} (Brookite), \cite{Columbite_Chen_2002-x} (Columbite) and \cite{Rutile_Swope_1995-x} (Rutile).}
\end{figure}

We find that PBEsol yields excellent equilibrium lattice parameters for known TiO$_2$ polymorphs (anatase, brookite, columbite, rutile), as shown in Fig.~\ref{fig:Latt_params_TiO2}. PBEsol also gives good relative static enthalpies between TiO$_2$ polymorphs, when benchmarked against accurate Diffusion Monte Carlo results \cite{Trail_PRB_2017-x}, and as discussed in the main text, correctly reproduces the structure and stability of known titanium carbides and oxides, as well as known aluminium carbides and oxides.


\newpage

\section{Phonon dispersion relations for titanium and aluminium carbonate} 
Calculations of phonon frequencies use density-functional perturbation theory \cite{CASTEP-DFPT-x} and implemented in version 18.1 of the CASTEP planewave pseudopotential code \cite{CASTEP-x}. We use the PBEsol exchange-correlation functional \cite{PBEsol-x}, in-built norm-conserving NCP pseudopotentials for Al, Ti, C and O, a planewave basis set cutoff of 1600 eV, and a Monkhorst-Pack \cite{MP_grid-x} Brillouin-zone sampling density of $<2\pi\times0.03$ \AA$^{-1}$.

\begin{figure*}[htp]
  \centering
  \subfigure{\includegraphics[width=8.7cm,clip]{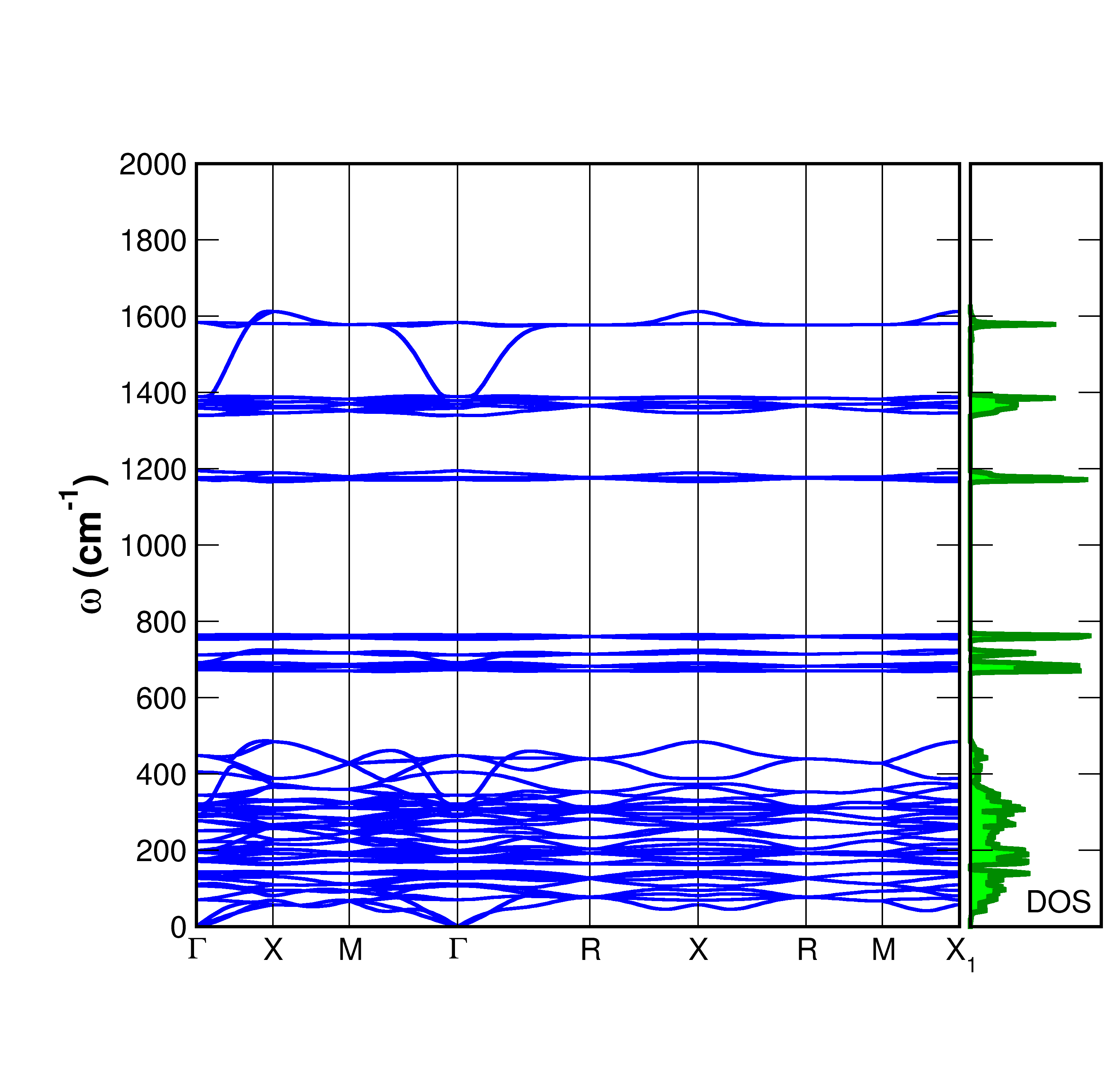}}\quad
  \subfigure{\includegraphics[width=8.7cm,clip]{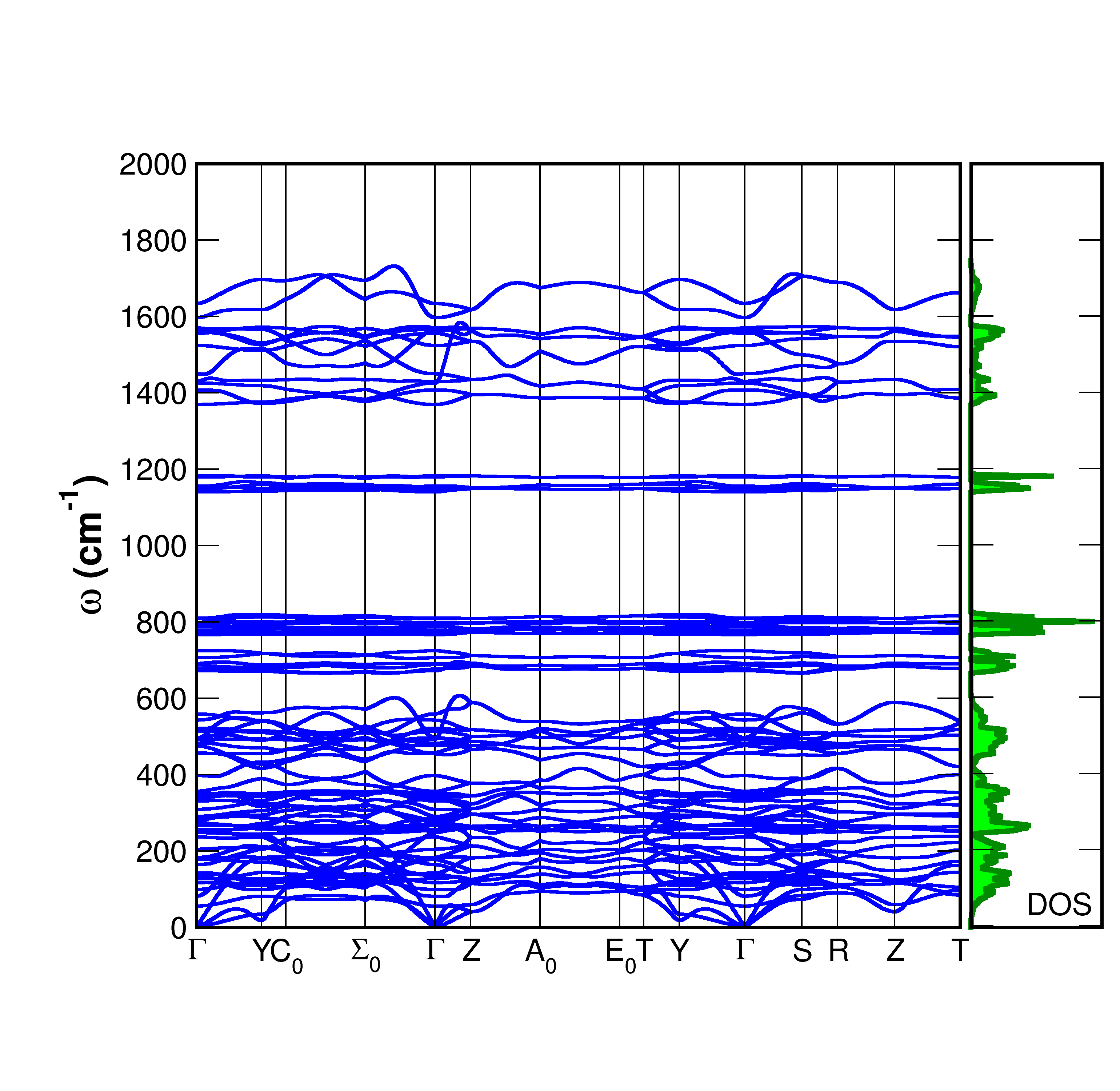}}
  \put(-385,215){Ti(CO$_3$)$_2$}
  \put(-130,215){Al$_2$(CO$_3$)$_3$}
  \caption{\label{fig:carbonate_phonons} (\textit{Left}) Phonon dispersion relations for titanium carbonate, space group $Pa3$, Ti(CO$_3$)$_2$. The RHS of the figure shows the phonon density of states (DOS). (\textit{Right}) Phonon dispersion relations for aluminium carbonate, space group $Cmcm$, Al$_2$(CO$_3$)$_3$. The RHS of the figure shows the phonon density of states (DOS). In both dispersion plots, the SeeK-path tool \cite{Seekpath-1-x,Seekpath-2-x,Seekpath-3-x} was used to calculate the Brillouin zone path.}
\end{figure*}

\section{Enthalpies and pressure tunabilities} 
In Fig.~4 of the main article, we consider oxide/carbonate pairs for the reaction MO+CO$_2$$\longrightarrow$MCO$_3$. DFT enthalpies are used to compute $\Delta H$ for this reaction. Here, we give various structures used for the metal oxides (MO), CO$_2$, and metal carbonates (MCO$_3$).

In all cases, the experimentally-known cubic $Pa3$ `Phase I' structure is used for solid CO$_2$.

For M = Ca, we have CaO in the rocksalt $Fm\overline{3}m$ structure (\#51409 in the ICSD). We consider CaCO$_3$ in the calcite and aragonite structures (\#18164 and \#15194), as well as 5 candidate structures for the vaterite phase, as given in Tables 1-5 in the Supplementary Information of Ref.~\cite{Demichelis_CrystEngComm_2012-x}.

For M = K, we have K$_2$O in the fluorite structure (\#44674), and we consider three structures for K$_2$CO$_3$ with space groups $C2/c$, $P2_1/c$ and $P6_3/mmc$, ICSD nos. 662, 10191 and 52535, respectively.

For M = Mg, we have MgO in the rocksalt $Fm\overline{3}m$ structure (\#9863 in the ICSD), and we consider MgCO$_3$ in the calcite structure (\#10264). 

For M = Na, we have Na$_2$O in the fluorite structure (\#60435), and we consider two structures for Na$_2$CO$_3$ with space groups $C2/m$ and $P6_3/mmc$, ICSD nos. 12168 and 81004, respectively.

For M = Si, we consider SiO$_2$ in the $\alpha$-cristobalite, $\alpha$-quartz, $\beta$-cristobalite, and $\beta$-quartz structures. We use the $P2_1/c$ structure determined for SiC$_2$O$_6$ (i.e.~Si(CO$_3$)$_2$) using evolutionary structure searches in Ref.~\cite{Marques_PCCP_2015-x}.

Finally for M = Ti and Al, we have TiO$_2$ in the anatase structure, space group $I4_1/amd$, and Al$_2$O$_3$ in its corundum structure, space group $R\overline{3}c$. The structures of their respective carbonates are as depicted in Fig.~3 of the main article.

\section*{Supplemental Material references}

\end{document}